\documentclass[12pt]{article}
\begin{document}
\begin{center}
{\large\bf Late-time Inhomogeneity and Acceleration Without Dark
Energy } \vskip 0.3 true in {\large J. W. Moffat} \vskip 0.3 true
in {\it The Perimeter Institute for Theoretical Physics, Waterloo,
Ontario, N2J 2W9, Canada} \vskip 0.3 true in and \vskip 0.3 true
in {\it Department of Physics, University of Waterloo, Waterloo,
Ontario N2L 3G1, Canada}
\end{center}
\begin{abstract}%
The inhomogeneous distribution of matter in the non-linear regime
of galaxies, clusters of galaxies and voids is described by an
exact, spherically symmetric inhomogeneous solution of Einstein's
gravitational field equations, corresponding to an under-dense
void. The solution becomes the homogeneous and isotropic
Einstein-de Sitter solution for a red shift $z > 10-20$, which
describes the matter dominated CMB data with small inhomogeneities
$\delta\rho/\rho\sim 10^{-5}$. A spatial volume averaging of
physical quantities is introduced and the averaged time evolution
expansion parameter $\theta$ in the Raychaudhuri equation can give
rise in the late-time universe to a volume averaged deceleration
parameter $\langle q\rangle$ that is negative for a positive
matter density. This allows for a region of accelerated expansion
which does not require a negative pressure dark energy or a
cosmological constant. A negative deceleration parameter can be
derived by this volume averaging procedure from the
Lema\^{i}tre-Tolman-Bondi open void solution, which describes the
late-time non-linear regime associated with galaxies and
under-dense voids and solves the ``coincidence'' problem.
\end{abstract}
\vskip 0.2 true in e-mail: john.moffat@utoronto.ca


\section{Introduction}

In a recent article, we investigated a cosmology in which a
spherically symmetric inhomogeneous enhancement is embedded in an
asymptotic FLRW universe~\cite{Moffat}. The inhomogeneous
enhancement is described by an exact inhomogeneous solution of
Einstein's field equations. We found that the inhomogeneities can
lead to a reinterpretation of the luminosity distance $d_L$ of a
cosmological source in terms of its red shift $z$, owing to the
observer dependence of these quantities. The time evolution and
the expansion rate of the inhomogeneous distribution of matter can
lead to intrinsic effects such as cosmic variance at large angles
and long-wavelength perturbations not described by a
Friedmann-Lema\^{i}tre-Robertson-Walker (FLRW) homogeneous and
isotropic universe. Therefore, the interpretation of the data
using a FLRW model that the accelerating expansion of the universe
is caused by dark energy may be misleading. This is important, for
it is difficult to explain theoretically the postulated dark
energy that causes the acceleration of the universe. The model
also leads to an axis pointing towards the center of the
spherically symmetric large scale inhomogeneous enhancement with
dipole, quadrupole and octopole moments aligned with the axis. It
was shown that the luminosity distances and red shifts observed by
different observers located at spatially different points of
causally disconnected parts of the universe can have varying
values. A spatial average of all these observations leads to an
intrinsic cosmic variance in e.g. the deceleration parameter $q$.
The distribution of CMB temperature fluctuations can be unevenly
distributed in the northern and southern hemispheres.

The acceleration of the expansion of the universe deduced from
Type SNIa supernova observations and the CMB WMAP
data~\cite{Perlmutter,Riess,Spergel} has been interpreted as due
to the cosmological constant (vacuum energy), modifications of
Einstein's gravitational field equations at large
distances~\cite{Turner}, and quintessence fields~\cite{Peebles}.
The quintessence explanations postulate a new form of matter with
negative pressure called dark energy. Recently, it has been
suggested that the acceleration is caused by very long wavelength,
super-horizon perturbations generated by a period of inflation in
the early universe~\cite{Kolb,Barausse,Notari,Rasanen}. The
back-reaction of perturbations on a FLRW background universe has
been the subject of investigation by several
authors~\cite{Brandenberger}. The predictions based on
perturbation theory are limited by the condition $\Phi\ll 1$,
where $\Phi$ is the gravitational potential and the assumption of
a homogeneous and isotropic FLRW background. Moreover, the
perturbation calculations do not possess a rigorous gauge
invariance unless all physical quantities are scalars under
coordinate transformations. Otherwise, calculations rely on using
specific gauges such as the synchronous gauge or the Poisson
(Newtonian) gauge, which can possess residual gauge ambiguities.

The suggestion that the acceleration of the universe can be caused
by sub-horizon or super-horizon modes due to a phase of early
universe inflation has been
criticized~\cite{Chung,Flanagan,Seljak,Wiltshire,Siegel,Giovannini}.
It is pointed out that second order perturbation effects of the
form $\Phi\nabla^2\Phi$ are described by a renormalization of the
local spatial curvature and cannot (for a positive energy density
and pressure) produce a negative deceleration parameter. Hitara
and Seljak~\cite{Seljak} use a derivation of the deceleration
parameter $q$ obtained from the
Raychaudhuri~\cite{Raychaudhuri,Hawking} equation that requires
significant contributions from vorticity to produce a negative $q$
when the strong energy condition $\rho+3p>0$ is satisfied.
Notari~\cite{Notari} argues that the perturbation expansion fails
at late times and leads to significant infrared effects and a
possible negative deceleration parameter $q$. All of these
arguments depend on a perturbation calculation against a
homogeneous and isotropic FLRW background. Thus, the results are
inferred from the strong assumption that the universe is
approximately homogeneous at all scales. We know that for
distances of order $\leq 150$ Mpc there are galaxies, clusters of
galaxies and large scale voids $\sim 50$ Mpc, and that
$\delta\rho/\rho\geq 1$, so that linear perturbation theory fails.

The inhomogeneous distribution of matter at late times could
possibly allow for a negative deceleration parameter $q$,
depending on the nature of the exact inhomogeneous solution of
Einstein's field equations. This will allow for the possibility of
explaining the acceleration of the universe, without a negative
pressure dark energy fluid or a cosmological constant. It can
solve the ``coincidence'' problem, because the negative
deceleration parameter $q$ and the accelerating expansion will not
occur in the absence of large-scale structure in the form of
galaxies, voids and clusters of galaxies \footnote{The possibility
that a large-scale inhomogeneity can mimic a cosmological constant
has been considered previously by C\'el\'erier~\cite{Celerier}.
The possibility that an under-dense void can cause the
acceleration of the universe has been proposed by
Tomita~\cite{Tomita}.}

We shall describe the inhomogeneous late-time universe by an exact
spherically symmetric solution of Einstein's field equations. The
solution becomes the homogeneous and isotropic Einstein-de Sitter
matter dominated solution for $z> 20$, so that it can describe the
WMAP data near the surface of last scattering with small
temperature perturbations $\delta T/T\sim 10^{-5}$. However, for
$z < 20$ the solution is inhomogeneous and for luminosity
distances corresponding to $z\sim 0.2-1.4$, includes the Type SNIa
supernova measurements. We investigate the expansion geometry of
the spherical distribution of inhomogeneous matter and under-dense
voids~\cite{Hoyle} to see whether it allows for the possibility of
a negative $q$ and an accelerating expansion of the universe,
without negative pressure dark energy or a cosmological constant.

A generic relativistic solution of a three-dimensional
inhomogeneous universe is presently unknown. The highly symmetric
Lema\^{i}tre-Tolman-Bondi (LTB)~\cite{Lemaitre,Tolman,Bondi}
solution that we shall consider in the following is inhomogeneous
in only one spatial dimension. Even though the assumption of
spherical symmetry is unrealistic, the model is useful in
illustrating important {\it local} features of physical
quantities.

We will discover that although the LTB model appears to allow for
a negative deceleration parameter for a positive density $\rho >
0$, an investigation of the gauge invariant Raychoudhuri equation
that determines the rate of change of the volume expansion shows
that for the synchronous and comoving LTB metric, the vorticity
$\omega$ vanishes and $q$ must be positive when the strong energy
condition is satisfied. However, for our inhomogeneous late-time
universe model, we are required to carry out a volume averaging of
physical scalar quantities, such as the expansion parameter
$\theta$. We find for such an averaging process that for a zero
cosmological constant and for irrotational matter the average
deceleration parameter $\langle q\rangle$ can be negative and
describe an accelerating local region of the universe,
corresponding to our observed Hubble radius, without invoking dark
energy with a negative pressure.

\section{Inhomogeneous Friedmann Equations}

For the sake of notational clarity, we write the FLRW line element
\begin{equation}
ds^2=dt^2-a^2(t)\biggl(\frac{dr^2}{1-kr^2}+r^2d\Omega^2\biggr),
\end{equation}
where $k=+1,0,-1$ for a closed, flat and open universe,
respectively, and $d\Omega^2=d\theta^2+\sin\theta^2d\phi^2$. The
general, spherically symmetric inhomogeneous line element is given
by~\cite{Lemaitre,Tolman,Bondi,Bonnor,Moffat3,Moffat4,Krasinski,Moffat}:
\begin{equation}
\label{inhomometric} ds^2=dt^2-X^2(r,t)dr^2-R^2(r,t)d\Omega^2.
\end{equation}
The energy-momentum tensor ${T^\mu}_\nu$ takes the barytropic form
\begin{equation}
\label{energymomentum} {T^\mu}_\nu=(\rho+p)u^\mu u_\nu
-p{\delta^\mu}_\nu,
\end{equation}
where $u^\mu=dx^\mu/ds$ and, in general, the density
$\rho=\rho(r,t)$ and the pressure $p=p(r,t)$ depend on both $r$
and $t$. We have for comoving coordinates $u^0=1, u^i=0,\,
(i=1,2,3)$ and $g^{\mu\nu}u_\mu u_\nu=1$.

The Einstein gravitational equations are
\begin{equation}
\label{Einstein} G_{\mu\nu}+\Lambda g_{\mu\nu}=-8\pi GT_{\mu\nu},
\end{equation}
where $G_{\mu\nu}=R_{\mu\nu}-\frac{1}{2}g_{\mu\nu}{\cal R}$,
${\cal R}=g^{\mu\nu}R_{\mu\nu}$ and $\Lambda$ is the cosmological
constant. Solving the $G_{01}=0$ equation for the metric
(\ref{inhomometric}), we find that
\begin{equation}
X(r,t)=\frac{R'(r,t)}{f(r)},
\end{equation}
where $R'=\partial R/\partial r$ and $f(r)$ is an arbitrary
function of $r$.

We obtain the two generalized Friedmann equations~\cite{Moffat}:
\begin{equation}
\label{inhomoFriedmann} \frac{{\dot R}^2}{R^2}+2\frac{{\dot
R}'}{R'}\frac{{\dot R}}{R}+\frac{1}{R^2}(1-f^2)
-2\frac{ff'}{R'R}=8\pi G\rho+\Lambda,
\end{equation}
\begin{equation}
\label{inhomoFriedmann2} \frac{\ddot
R}{R}+\frac{1}{3}\frac{\dot{R}^2}{R^2}
+\frac{1}{3}\frac{1}{R^2}(1-f^2) -\frac{1}{3}\frac{{\dot
R}'}{R'}\frac{{\dot R}}{R}+\frac{1}{3} \frac{ff'}{R'R}
=-\frac{4\pi G}{3}(\rho+3p)+\frac{1}{3}\Lambda,
\end{equation}
where $\dot R=\partial R/\partial t$.

\section{Late-Time Matter Dominated Universe}

The late-time matter dominated universe will be pictured as a
large-scale inhomogeneous enhancement that is described by an
exact inhomogeneous spherically symmetric solution of Einstein's
field equations. The inhomogeneous enhancement is embedded in a
matter dominated universe that approaches asymptotically an
Einstein-de Sitter universe as $t\rightarrow\infty$. An observer
will be off-center from the origin of coordinates of the spherical
inhomogeneous enhancement. The local inhomogeneous solution
corresponds to an open, hyperbolic solution and describes an
under-dense void. Surveys such as the 2-degree Field Galaxy
Redshift Survey and the Sloan Digital Sky Survey show large
volumes of relatively empty space, or voids, in the distribution
of galaxies~\cite{Hoyle}.

For the matter dominated Lema\^{i}tre-Tolman-Bondi
(LTB)~\cite{Lemaitre,Tolman,Bondi} model with zero pressure $p=0$
and zero cosmological constant $\Lambda=0$, the Einstein field
equations demand that $R(r,t)$ satisfies
\begin{equation}
\label{Requation} 2R(r,t){\dot R}^2(r,t)+2R(r,t)(1-f^2(r))=F(r),
\end{equation}
with $F$ being an arbitrary function of $r$ of class $C^2$.

The proper density of matter can be expressed as
\begin{equation}
\label{density} \rho=\frac{F'}{16\pi GR'R^2}.
\end{equation}
We can solve (\ref{density}) to obtain
\begin{equation}
\Omega-1\equiv\frac{\rho}{\rho_c}-1=\frac{1}{H^2_{\rm
eff}}\biggl(\frac{1-f^2}{R^2}-2\frac{f}{R}\frac{f'}{R'}\biggr),
\end{equation}
where
\begin{equation}
H^2_{\rm eff}=H^2_\perp+2H_\perp H_r,
\end{equation}
is an effective Hubble parameter and
\begin{equation}
H_r=\frac{{\dot R}'}{R'},\quad H_\perp=\frac{{\dot R}}{R}.
\end{equation}
We have for the critical density
\begin{equation}
8\pi G\rho_c=H^2_{\rm eff}.
\end{equation}
There are three possibilities for the curvature of spacetime: 1)
$f^2 > 1$ open ($\Omega-1 < 0$), 2) $f^2=1$ flat ($\Omega-1=0$),
$f^2 < 1$ closed ($\Omega-1 > 0$).

We describe the inhomogeneous density regime by a hyperbolic, open
solution $f^2(r) > 1$ with $p=\Lambda=0$, corresponding to an
under-dense void. We choose
\begin{equation}
f(r)=\sqrt{1+r^2}
\end{equation}
and the metric reduces to
\begin{equation}
\label{openmetric}
ds^2=dt^2-\frac{R^{'2}(r,t)}{1+r^2}dr^2-R^2(r,t)d\Omega^2.
\end{equation}
A parametric solution is given by
\begin{equation}
\label{inhomoopen} R(r,t)=\frac{1}{4}F(r)(f^2(r)-1)^{-1}(\cosh
u(r,t)-1),
\end{equation}
\begin{equation}
t+\beta(r)=\frac{1}{4}(f^2(r)-1)^{-3/2}(\sinh u(r,t)-u(r,t)),
\end{equation}
\begin{equation}
8\pi G\rho(r,t)
=\biggl(\frac{2F'(r)}{F^2(r)}\biggr)\biggl(\frac{f^2(r)-1}{R'(r,t)}\biggr)
\biggl(\frac{1}{\sinh^4\frac{1}{2}u(r,t)}\biggr).
\end{equation}
Here, $\beta(r)$ is an arbitrary function of $r$ of class
$C^2$~\cite{Bonnor}.

The function $\beta(r)$ can be specified in terms of a density on
some spacelike hypersurface $t=t_0$. The metric and density are
singular for
\begin{equation}
u=0\quad {\rm or}\quad R'(r,t)=0,
\end{equation}
where $u=0$ refers to the hypersurface $t+\beta(r)=0$ and the
singular surface for $R'(r,t)=0$ is more complicated. Our
pressureless model requires that the singular surface
$t(r)=\Sigma(r)$ describes the surface on which the universe
becomes matter dominated (in the FLRW model this occurs at $z\sim
10^4$). The hyperbolic FLRW model is obtained by choosing the
conditions
\begin{equation}
f^2(r)=1+r^2,\quad F(r)=4sr^3,\quad \beta(r)=0,
\end{equation}
where $s$ is a positive constant. This leads to the metric
solution
\begin{equation}
R=sr(\cosh u-1),
\end{equation}
\begin{equation}
t=s(\sinh u-u),
\end{equation}
\begin{equation}
8\pi
G\rho=\frac{3}{4s^2}\biggl(\frac{1}{\sinh^6\frac{1}{2}u}\biggr).
\end{equation}

We shall change the radial coordinate to
\begin{equation}
r^*=r-r_{SLS},
\end{equation}
where $r=r_{SLS}$ is the location of the surface of last
scattering corresponding to $r^*=0$ at a red shift $z\sim 1100$.
We postulate that $\beta(r^*)$ approaches a maximum constant value
at the approximate end of galaxy and cluster formation, $z_f\sim
1-5$, where $z=z_f$ denotes the final value of the red shift. We
assume that $\beta(r^*)\rightarrow 0$ and $f(r^*)\rightarrow 1$ as
$r^*\rightarrow 0$. A model for $\beta$ is of the form:
\begin{equation}
\beta(r^*)=a\biggl(\frac{r^*}{b+r^*}\biggr)^\alpha,
\end{equation}
where $a$, $b$ and $\alpha$ are positive constants, $\beta(0)=0$
and $\beta(r^*)\rightarrow a$ for $r^* \gg b$. The surface $r^*=0$
for the metric is a singular surface.

The metric for $f(r^*)\rightarrow 1$ becomes~\cite{Bonnor}:
\begin{equation}
\label{matterdommetric}
ds^2=dt^2-(t+\beta(r^*))^{4/3}(Y^2(r^*,t)dr^{*2}+r^{*2}d\Omega^2),
\end{equation}
where
\begin{equation}
Y(r^*,t)=1+\frac{2r^*\beta'(r^*)}{3(t+\beta(r^*))},
\end{equation}
and
\begin{equation}
\rho(r^*,t)=\frac{1}{6\pi G(t+\beta(r^*))^2Y(r^*,t)}.
\end{equation}
We obtain for $\beta(r^*)\rightarrow 0$ the homogeneous and
isotropic solution corresponding to a spatially flat Einstein-de
Sitter universe with the metric
\begin{equation}
ds^2=dt^2-a^2(t)(dr^{*2}+r^{*2}d\Omega^2),
\end{equation}
and $a(t)\propto t^{2/3}$. This solution is compatible as a
background spacetime with the approximately homogeneous and
isotropic WMAP data for $20 < z < 10^3$ and $\delta\rho/\rho\sim
10^{-5}$~\cite{Spergel}.

The Einstein-de Sitter solution must be supplemented by non-zero
pressure contributions for $z\sim 10^3$ in the linear regime at
the surface of last scattering in order to take into account the
radiation density $\rho_R$. Moreover, a more general solution is
required that has a non-vanishing cosmological constant $\Lambda$,
so that we have $\Omega=\Omega_M+\Omega_R +\Omega_\Lambda=1$ in
agreement with the spatially flat result $\Omega=1.02\pm 0.02$,
obtained from the determination of the first acoustic peak in the
WMAP power spectrum~\cite{Spergel}. However, in the following
section, we shall replace $\Omega_\Lambda$ by a contribution
corresponding to an acceleration parameter $q$ obtained from our
inhomogeneous void solution with $\Lambda=0$.

\section{The Inhomogeneous Late-Time Density and Deceleration Parameters}

We obtain from (\ref{inhomoFriedmann}) for $f(r)=\sqrt{1+r^2}$ and
$p=0$ (we replace for convenience $r^*$ by $r$):
\begin{equation}
H_\perp^2+2H_rH_\perp-\frac{r^2}{R^2}-\frac{2r}{R'R}=8\pi
G\rho+\Lambda.
\end{equation}
Dividing this equation by $H_\perp^2$ gives
\begin{equation}
\label{omegadensity} \Omega\equiv\frac{8\pi
G\rho}{H_\perp^2}+\frac{\Lambda}{H^2_\perp}+\frac{r^2}{R^2H_\perp^2}
+\frac{2r}{R'RH_\perp^2}-\frac{2H_r}{H_\perp}=1.
\end{equation}

Let us expand $R(r,t)$ in a Taylor series
\begin{equation}
R(r,t)=R[r,t_0-(t_0-t)] =R(r,t_0)\biggl[1-(t_0-t)\frac{{\dot
R}(r,t_0)}{R(r,t_0)} +\frac{1}{2}(t_0-t)^2\frac{{\ddot
R}(r,t_0)}{R(r,t_0)} -...\biggr]
$$ $$
=R(r,t_0)\biggl[1-(t_0-t)H_{0\perp}-\frac{1}{2}(t_0-t)^2q(r,t_0)H_{0\perp}^2-...\biggr],
\end{equation}
where $t_0$ denotes the present epoch and $H_{0\perp}={\dot
R}(r,t_0)/R(r,t_0)$. Moreover, we have
\begin{equation}
q(r,t_0)=-\frac{{\ddot R}(r,t_0)R(r,t_0)}{{\dot R}^2(r,t_0)}.
\end{equation}
For the matter dominated era we set $p=0$ in
Eqs.(\ref{inhomoFriedmann}) and (\ref{inhomoFriedmann2}) and
substitute for ${\ddot R}$ from (\ref{inhomoFriedmann2}) to obtain
\begin{equation}
\label{inhomodeceleration}
q(r,t_0)=\frac{1}{3}+\frac{4\pi\rho_0(r,t_0)}{3H_{0\perp}^2(r,t_0)}
$$ $$
+\frac{1}{3}\frac{r}{RR'H^2_{0\perp}}
-\frac{\Lambda}{3H^2_{0\perp}(r,t_0)}-\frac{1}{3}\frac{H_{0r}(r,t_0)}{H_{0\perp}(r,t_0)}
-\frac{1}{3}\frac{r^2}{R^2H^2_{0\perp}},
\end{equation}
where $H_{0r}(r,t_0)={\dot R}'(r,t_0)/R'(r,t_0)$.

If we set $H_{0\perp}(r,t_0)=H_{0r}(r,t_0)=H(t_0)$ where
$H(t_0)={\dot a}(t_0)/a(t_0)$ and $R(r,t_0)=a(t_0)$, then we
obtain the {\it global} FLRW expression for the deceleration
parameter for a spatially flat universe with $f^2(r)=1$:
\begin{equation}
q_0=\frac{1}{2}\Omega_{0M}-\Omega_{0\Lambda},
\end{equation}
where from (\ref{omegadensity}) $\Omega_{0M}+\Omega_{0\Lambda}=1$
with $\Omega_{0M}=8\pi\rho_{0M}/3H^2(t_0)$ and
$\Omega_{0\Lambda}=\Lambda/3H^2(t_0)$.

The variance of $q$ is given by the exact non-perturbative
expression:
\begin{equation}
{\rm var}(q)\equiv\langle q^2-\langle
q\rangle^2\rangle^{1/2}=\overline{q},
\end{equation}
where
\begin{equation}
\label{spatialaverage} \overline{(...)}=\frac{\int
d^3x\sqrt{\gamma}(...)}{\int d^3x\sqrt{\gamma}},
\end{equation}
denotes the ensemble average and $\gamma$ denotes the determinant
of the 3-dimensional metric $g_{ij}\,(i=1,2,3)$.

We see from (\ref{inhomodeceleration}) that different observers
located in different causally disconnected parts of the sky will
observe different values for the deceleration parameter $q$,
depending upon their location and distance from the center of the
spherically symmetric inhomogeneous enhancement. This can lead to
one form of cosmic variance, because the spatial average of all
the observed values of local physical quantities, including the
deceleration parameter, $q$, will have an intrinsic uncertainty.

Let us set the cosmological constant to zero, $\Lambda=0$
\footnote{We do not provide any solution to the cosmological
constant problem, namely, why $\Lambda=0$.}. From
(\ref{omegadensity}) and (\ref{inhomodeceleration}), we obtain for
$t=t_0$:
\begin{equation}
\label{density2} \Omega_0\equiv \frac{8\pi
G\rho_0}{H^2_{0\perp}}+\frac{r^2}{R^2H^2_{0\perp}}
+\frac{2r}{R'RH^2_{0\perp}}-\frac{2H_{0r}}{H_{0\perp}}=1,
\end{equation}
and
\begin{equation}
\label{inhomodeceleration2} q(r,t_0)=\frac{1}{3} +\frac{4\pi
G\rho_0(r,t_0)}{3H^2_{0\perp}(r,t_0)}
+\frac{1}{3}\frac{r}{RR'H^2_{0\perp}}
-\frac{1}{3}\frac{H_{0r}(r,t_0)}{H_{0\perp}(r,t_0)}-\frac{1}{3}\frac{r^2}{R^2H^2_{0\perp}}.
\end{equation}

If in (\ref{inhomodeceleration2}) we have
\begin{equation}
\label{accelcond} H_{0r}H_{0\perp}+\frac{r^2}{R^2} > 4\pi
G\rho_0+H^2_{0\perp}+\frac{r}{RR'},
\end{equation}
then the deceleration parameter $q$ can be negative {\it and cause
the universe to accelerate without a cosmological constant or dark
energy}. If we solve for $H_{0r}/H_{0\perp}$ from
Eq.(\ref{density2}) and substitute into
(\ref{inhomodeceleration2}), we obtain
\begin{equation}
\label{qacceleration}
q(r,t_0)=\frac{1}{2}-\frac{5}{6}\biggl(\frac{r^2}{R^2H^2_{0\perp}}\biggr),
\end{equation}
which gives a negative $q$ for
\begin{equation}
\label{accelcond2} r^2/R^2H^2_{0\perp}> 3/5,
\end{equation}
and we can satisfy the condition (\ref{density2}). We see that for
$r=0$ corresponding to $f^2(r)=1$ and a spatially flat universe
$q(r,t_0)=1/2$. For $p=\Lambda=0$, Eqs.(\ref{inhomoFriedmann2})
and (\ref{accelcond}) lead to ${\ddot R}> 0$ as the condition for
an accelerating expansion of the universe.

The conditions (\ref{density2}) and (\ref{accelcond}) require that
$f^2(r)> 1$ for the inhomogeneous enhancement, corresponding to an
under-dense void. If we choose instead to describe the
inhomogeneity by the spatially flat solution $f^2(r)=1$, then we
cannot simultaneously satisfy $\Omega=1$ and $q <0$. However, we
will investigate in the next Section, whether the condition
(\ref{accelcond}) can indeed be satisfied for the particular LTB
model that we have adopted. We shall find that a more general
inhomogeneous solution is needed to establish whether such a
solution can explain the accelerating expansion of the universe
without a negative pressure fluid or a cosmological constant.

Several authors have investigated the robustness of the distance
of the FLRW model determination of the magnitude red shift
relation and the determination of the luminosity distance $D_L$ as
a function of the parameters $z,H_0,\Omega_M$ and
$\Omega_\Lambda$~\cite{Mashhoon,Liang,Celerier}. In particular, an
analysis of the magnitude red shift relation by
C\'el\'erier~\cite{Celerier} reveals that there exists a
significant degeneracy and uncertainty in the determination of
these parameters when analyzed using a LTB model to describe a
local inhomogeneity.

\section{The Raychaudhuri Equation}

We must now investigate whether the conditions
Eqs.(\ref{accelcond}) and (\ref{accelcond2}) can be satisfied by
our LTB model. We will find that this is not the case and that we
have to consider more general inhomogeneous solutions. To this
end, we shall investigate the consequences of the Raychaudhuri
equation~\cite{Raychaudhuri,Hawking}, which holds in general for
any cosmological solution based on Einstein's gravitational field
equations.

We shall consider a congruence of curves with a time-like unit
vector $V_\mu$ with $g^{\mu\nu}V_\mu V_\nu=1$ and
$dV^\mu/ds=\nabla_\nu V^\mu V^\nu$ is the acceleration of the flow
lines. The metric tensor $h^{\mu\nu}$ is given by
\begin{equation}
{h^\mu}_\nu={\delta^\mu}_\nu+V^\mu V_\nu
\end{equation}
and describes the metric that projects a vector into its
components in the subspace of the vector tangent space that is
orthogonal to $V$.

We define the vorticity tensor
\begin{equation}
\omega_{\mu\nu}=\nabla_{[\nu} V_{\mu]},
\end{equation}
and the shear tensor
\begin{equation}
\sigma_{\mu\nu}=\theta_{\mu\nu}-\frac{1}{3}\theta h_{\mu\nu},
\end{equation}
where
\begin{equation}
\theta_{\mu\nu}=\nabla_{(\nu}V_{\mu)}.
\end{equation}
The covariant derivative of $V$ can be expressed as
\begin{equation}
\nabla_\nu
V_\mu=\omega_{\mu\nu}+\sigma_{\mu\nu}+\frac{1}{3}h_{\mu\nu}\theta
-\frac{dV_\mu}{ds}V_\nu.
\end{equation}
The volume expansion is given by
\begin{equation}
\label{volumeexpand} \theta\equiv
h^{\mu\nu}\theta_{\mu\nu}=\nabla_\mu V^\mu.
\end{equation}

The Raychaudhuri equation is
\begin{equation}
\label{Ray} \frac{d\theta}{ds}=-R_{\mu\nu}V^\mu
V^\nu+2\omega^2-2\sigma^2-\frac{1}{3}\theta^2,
\end{equation}
where we have adopted geodesic world lines and
\begin{equation}
\omega^2=\omega^{\mu\nu}\omega_{\mu\nu} \geq 0,\quad
\sigma^2=\sigma^{\mu\nu}\sigma_{\mu\nu}\geq 0.
\end{equation}
From Einstein's field equations, we have for $p=\Lambda=0$:
\begin{equation}
\label{Ray2}
\frac{d\theta}{ds}=2\omega^2-2\sigma^2-\frac{1}{3}\theta^2-4\pi
G\rho.
\end{equation}
The deceleration parameter $q$ is defined by
\begin{equation}
q\equiv
-\frac{(3d\theta/ds+\theta^2)}{\theta^2}=\frac{6(\sigma^2-\omega^2)+4\pi
G\rho} {\theta^2}.
\end{equation}
If the vorticity $\omega=0$, then the deceleration parameter $q$
has to be positive for positive $\rho$.

Let us now consider the inhomogeneous metric (\ref{inhomometric})
and the generalized Friedmann equations (\ref{inhomoFriedmann})
and (\ref{inhomoFriedmann2}). The metric is chosen to satisfy the
synchronous gauge and comoving conditions $g_{00}=1$ and
$g_{0i}=0$ ($i=1,2,3$) corresponding to $V^\mu=(1,0,0,0)$. We have
\begin{equation}
\nabla_\mu V^\mu=\partial_\mu V^\mu+{\Gamma^\sigma}_{\alpha\sigma}
V^\alpha.
\end{equation}
We obtain for the metric (\ref{inhomometric}):
\begin{equation}
\nabla_\mu V^\mu={\Gamma^\sigma}_{0\sigma}={\Gamma^i}_{0i}.
\end{equation}
This in turn becomes (see Appendix A in~\cite{Moffat}):
\begin{equation}
\label{inhomotheta} \theta=\frac{{\dot R}'}{R'}+2\frac{{\dot
R}}{R}=H_r+2H_\perp.
\end{equation}

For our synchronous comoving coordinates we have
\begin{equation}
\frac{d\theta}{ds}=V^\mu\nabla_\mu\theta=V^\mu\partial_\mu\theta={\dot\theta}.
\end{equation}
From (\ref{inhomotheta}) we get
\begin{equation}
q\equiv
-\frac{(3\dot\theta+\theta^2)}{\theta^2}=-\frac{\biggl(3\frac{{\ddot
R}'}{R'}+6\frac{{\ddot
R}}{R}+\theta^2-3H_r^2-6H_\perp^2\biggr)}{\theta^2}.
\end{equation}
We may now conclude that if
\begin{equation}
\label{qaccel} 3H_r^2+6H_\perp^2 > 3\frac{{\ddot
R}'}{R'}+6\frac{{\ddot R}}{R}+\theta^2,
\end{equation}
then $q$ could be negative leading to a local acceleration of the
universe. However, for our synchronous comoving frame, we have
\begin{equation}
\omega_{\mu\nu}\equiv \nabla_{[\nu}V_{\mu]}=\partial_\nu
V_\mu-\partial_\mu V_\nu=0,
\end{equation}
so that the vorticity for this chosen gauge is zero. Therefore, we
must deduce that the condition (\ref{qaccel}) cannot be satisfied
for the positive energy condition $\rho > 0$. Moreover, the two
conditions (\ref{accelcond}) and (\ref{accelcond2}) cannot be
satisfied for the synchronous and comoving metric
(\ref{inhomometric}). However, in the following section, we will
show that we must carry out a spatial volume averaging of the
expansion parameter $\theta$ and its time evolution to arrive at a
physically viable description of the local inhomogeneous late-time
universe and an accelerating expansion of the universe.

\section{Spatially Averaged Cosmological Domains}

For our inhomogeneous model, it is necessary to perform a spatial
average (\ref{spatialaverage}) of physical quantities, due to
their observer, location dependence. If we postulate that the
universe satisfies the cosmological principle and spacetime is
described by the homogeneous and isotropic FLRW metric, then we
can make {\it global} predictions about the universe within our
Hubble horizon. However, in general we cannot assume that the
universe is homogeneous and isotropic, so we determine the
structure of spacetime from the cosmological data available in our
past light cone. By adopting a ``smoothing'' or averaging of the
granular structure within a domain ${\cal D}$ in the past light
cone, we can infer the properties of spacetime using this
averaging process outside our past light
cone~\cite{Ellis,Buchert,Kolb2}.

Although the gravitational potential $\Phi$ is weak, $\Phi\ll 1$,
we have to treat any back-reaction effect due to inhomogeneities
in a relativistic way accounting for the non-linearity of
Einstein's field equations. Sizable post-Newtonian effects due to
an acceleration of the universe will appear in the Friedmann
equations, describing an inhomogeneous universe after the
smoothing or averaging process has been carried out. After a
choice of time slicing is made to carry out the volume spatial
averaging over a domain ${\cal D}$, the question whether an
observer sees the universe accelerate or not depends on the data
set the observer uses and the size of the scale over which the
averaging process is invoked. Our model for the accelerating
universe is based on a local void which plays the role of a
cosmological constant. An observer living near or inside the void
sees the local region of spacetime expand faster than the outer
region of the void, despite the fact that both regions are
decelerating. Thus, even though individual local fluid elements
experience a deceleration, the averaged domain including the void
and its outer regions will appear to the observer as an
accelerating universe.

An open question regarding the averaging procedure is whether it
is covariant. The questions of covariance, gauge invariance and
the uniqueness of the averaging procedure remain to be solved. The
Einstein field equations are non-linear, so that the average of
the tensor density of matter does not commute with the average of
the time derivatives of the metric tensor occurring in the
Einstein tensor. The average of the expansion of the universe
derived from the luminosity distance must be consistently matched
with the results obtained from the volume expansion parameter
$\theta$ and the average of the density of matter.

Let us define a  more specific domain averaged expression for a
scalar quantity~\cite{Ellis,Buchert,Kolb2}:
\begin{equation}
\langle\Psi({\vec x},t)\rangle_D=\frac{1}{{\cal V_D}}\int_D
d^3x\sqrt{\gamma}\Psi({\vec x},t),
\end{equation}
where
\begin{equation}
{\cal V}_D=\int_Dd^3x\sqrt{\gamma}
\end{equation}
is the volume of the simply-connected domain, $D$, in a
hypersurface. We can define an effective scale-factor for our
spatially averaged spherically symmetric inhomogeneous enhancement
(void):
\begin{equation}
\langle R(r,t)\rangle_D=\biggl(\frac{{\cal V}(t)_D}{{\cal
V}_{iD}}\biggr)^{1/3},
\end{equation}
where ${\cal V}_{iD}$ is the initial volume.

The volume averaging of the scalar $\Psi$ does not commute with
its time evolution~\cite{Ellis,Buchert,Kolb2}:
\begin{equation}
\langle\dot{\Psi}(r,t)\rangle_D-\partial_t\langle\Psi(r,t)\rangle_D
=\langle\Psi(r,t)\rangle_D\langle\theta(r,t)\rangle_D
-\langle\Psi(r,t)\theta(r,t)\rangle_D.
\end{equation}
We can derive for the averaged $\langle\theta\rangle_D$ the
equation
\begin{equation}
\langle\theta\rangle_D=3\frac{\langle{\dot R}\rangle_D}{\langle
R\rangle_D}=3\langle H_{\perp}\rangle_D.
\end{equation}
Setting $\Psi=\theta$ and substituting for the Raychaudhuri
equation (\ref{Ray2}) for an irrotational matter dominated
late-time model with $\omega_{\mu\nu}=0$, $d\theta/ds=d\theta/dt$
and $\Lambda=0$, we obtain the expression
\begin{equation}
3\frac{\langle{\ddot R}\rangle_D}{\langle R\rangle_D}=-4\pi
G\langle\rho\rangle_D+\langle P\rangle_D,
\end{equation}
where $\langle P\rangle_D$ is a function of
$\langle\theta\rangle_D$, the average Hubble expansion parameters
$\langle H_\perp\rangle_D$ and $\langle H_r\rangle_D$ and the
average shear $\langle\sigma\rangle_D$.

For inhomogeneous cosmology, the smoothing due to averaging of the
Einstein field equations does not commute with the time evolution
of the non-linear field equations. This leads to extra
contributions in the effective, averaged Einstein field equations,
which do not satisfy the usual energy conditions even though they
are satisfied by the original energy-momentum tensor. It is the
lack of commutativity of the time evolution of the expansion of
the universe in a local patch inside our Hubble horizon that
circumvents the no-go theorem based on the local Raychaudhuri
equation.

If we can satisfy the condition
\begin{equation}
\label{Pcondition} \langle P\rangle_D >  4\pi
G\langle\rho\rangle_D,
\end{equation}
then the averaged deceleration parameter $\langle q\rangle_D$ can
be negative corresponding to the acceleration of a local patch of
the universe. If we substitute the value of $\langle{\ddot
R}\rangle_D$ from the volume averaged equation
(\ref{inhomoFriedmann2}) for $p=\Lambda=0$, then the exact LTB
solution for an {\it irrotational} inhomogeneous, under-dense void
solution can describe the local acceleration of the late-time
universe and fulfill the constraint equation (\ref{Pcondition}).
The spatial domain average of $q$ in Eq.(\ref{qacceleration})
gives
\begin{equation}
\label{qaccelerationaverage} \langle
q\rangle_D=\frac{1}{2}-\frac{5}{6}\left<\biggl(\frac{r^2}{R^2H^2_{0\perp}}\biggr)\right>_D.
\end{equation}
This yields the averaged condition for an accelerating universe:
\begin{equation}
\left<\frac{r^2}{R^2H^2_{0\perp}}\right>_D > 3/5.
\end{equation}
This spatial domain averaged condition for our LTB void solution
can be satisfied in our synchronous comoving gauge with zero
vorticity for $\langle\rho\rangle_D > 0$.

\section{Conclusions}

We have modelled the late-time inhomogeneous and non-linear regime
for $z < 10$ by an exact spherically symmetric, inhomogeneous
solution of Einstein's field equations. The solution approaches an
Einstein-de Sitter matter dominated solution for $z > 10$, which
describes a homogeneous and isotropic FLRW background spacetime
with small inhomogeneities $\delta\rho/\rho\sim 10^{-5}$ near the
surface of last scattering in agreement with the WMAP3
data~\cite{Spergel}. An investigation of the Raychaudhuri equation
for the expansion rate of the universe, shows that for the
synchronous and comoving gauge condition satisfied by the
inhomogeneous spherically symmetric metric, the vorticity $\omega$
is zero and the deceleration parameter $q$ is positive or zero
when the positive energy condition is satisfied. However, a volume
averaging of physical quantities is required for our inhomogeneous
universe, and due to the lack of commutativity of the time
evolution of the average of the scalar expansion $\theta$, we find
that it is possible to have a negative averaged deceleration
parameter $\langle q\rangle$ for a positive density of matter
$\rho$ and a zero vorticity $\omega=0$. Observational bounds on
the magnitude of the quantity $\langle P_D\rangle$ in the
condition (\ref{Pcondition}) must be obtained to establish whether
a sufficiently negative $q$ can be produced by the expanding
inhomogeneous enhancement or void to allow an explanation of the
accelerating universe without a negative pressure fluid or a
cosmological constant.

The model inhomogeneous solution is exact, so that we do not have
to be concerned with the failure of perturbative backreaction
calculations in the non-linear late-time universe regime. However,
we have assumed a high degree of symmetry for our inhomogeneous
solution, since it only allows for a one-dimensional inhomogeneity
in the radial direction, although the angular azimuthal expansion
described by $H_\perp$ does play a significant role in determining
the cosmological solution and the magnitude of the acceleration.
Due to the need for approximate isotropy, we require that the
observer be situated not too far from the ``center'' of the
spherically symmetric, inhomogeneous distribution of matter, so
this could be interpreted as a ``coincidence'' problem. But this
could be an artifact of the degree of symmetry assumed for the
model solution. It would be interesting to investigate
inhomogeneous cosmological solutions that possess less symmetry
than the LTB models~\cite{Krasinski}.

\vskip 0.2 true in {\bf Acknowledgments} \vskip 0.2 true in This
work was supported by the Natural Sciences and Engineering
Research Council of Canada. I thank Joel Brownstein and Martin
Green for helpful discussions.\vskip 0.5 true in

\end{document}